\documentclass[aps,prl,twocolumn,superscriptaddress]{revtex4-2}
\usepackage{graphicx}
\usepackage{color}

\usepackage{mhchem}

\begin{document}

\title{Line shapes in time- and angle-resolved photoemission spectroscopy explored by machine learning}

\author{Tami C. Meyer}
\affiliation {Department of Physics and Astronomy, Aarhus University, 8000 Aarhus C, Denmark}
\author{Gesa-R. Siemann}
\affiliation {Department of Physics and Astronomy, Aarhus University, 8000 Aarhus C, Denmark}
\author{Paulina Majchrzak}
\affiliation{Department of Applied Physics, Stanford University, Stanford, CA, USA}
\author{Thomas Seyller}
\affiliation{Institute of Physics, Chemnitz University of Technology, 09126 Chemnitz, Germany}
\affiliation{Center for Materials, Architectures and Integration of Nanomembranes (MAIN), Chemnitz University of Technology, 09126 Chemnitz, Germany}
\author{Jennifer Rigden}
\affiliation {Central Laser Facility, STFC Rutherford Appleton Laboratory, OX11 0QX, Harwell, UK}
\author{Yu Zhang}
\affiliation {Central Laser Facility, STFC Rutherford Appleton Laboratory, OX11 0QX, Harwell, UK}
\author{Emma Springate}
\affiliation {Central Laser Facility, STFC Rutherford Appleton Laboratory, OX11 0QX, Harwell, UK}
\author{Charlotte Sanders}
\affiliation {Central Laser Facility, STFC Rutherford Appleton Laboratory, OX11 0QX, Harwell, UK}
\author{Philip Hofmann}
\affiliation {Department of Physics and Astronomy, Aarhus University, 8000 Aarhus C, Denmark}

\begin{abstract}  
Time- and angle-resolved photoemission spectroscopy is a powerful technique for investigating the dynamics of excited carriers in quantum materials. Typically, data analysis proceeds via the inspection of time distribution curves (TDCs), which represent the time-dependent photoemission intensity in a region of interest---often chosen somewhat arbitrarily---in energy-momentum space. Here, we employ $k$-means, an unsupervised machine learning technique, to systematically investigate trends in TDC line shape for quasi-free-standing monolayer graphene and for a simple analytical model. Our analysis reveals how finite energy and time resolution can affect the TDC line shape. We discuss how this can be taken into account in a quantitative analysis, and under what conditions the time-dependent photoemission intensity after laser excitation can be approximated by a simple exponential decay.
\end{abstract}

\maketitle

\section{Introduction}

The properties of quantum materials are governed by the interactions of the electrons with various degrees of freedom, such as vibrations, charge and orbital dynamics. The strong coupling and intertwinement of these interactions often lead to the emergence of novel and exotic phases of matter. Disentangling these contributions is one of the major challenges in today's condensed matter physics. A promising avenue towards this goal is the study of quantum materials out-of-equilibrium, for instance as induced by a laser pulse \cite{Xu:2025aa}. In the simplest case, the excitation of a material by an ultrashort laser pulse leads to a hot electron population which decays with time, coupling hot electrons to different decay channels \cite{Perfetti:2007aa,Johannsen:2013ab}. In more complex scenarios, the excitation causes the destruction of certain orders, such as charge density waves or superconductivity, and the subsequent recovery can provide insight into the mechanism driving these orders \cite{Rohwer:2011aa,Hellmann:2012aa,Smallwood:2015aa,Zonno:2021ac}. Finally, there is the possibility of a light-induced generation of entirely new states that do not exist in equilibrium, such as superconductivity or Bloch-Floquet states  \cite{Duan:2023aa,Wang:2013ad,Budden:2021vz,Torre:2021wg}. 

Time- and angle-resolved photoemission spectroscopy (TR-ARPES) is a powerful technique for studying ultrafast dynamics in solids, as it provides direct access to the out-of-equilibrium spectral function \cite{Boschini:2024aa}.
TR-ARPES measures the photoemission intensity $I(\mathbf{k},\omega, t)$ as a function of momentum $\mathbf{k}$, energy $\omega$, and time $t$ as \begin{equation} 
I(\mathbf{k},\omega, t) \simeq A(\mathbf{k},\omega, t)\vert M(\mathbf{k},\omega,t) \vert^2 f(\omega,T_t)\ast G(\Delta \mathbf{k}, \Delta \omega, \Delta t), \label{eq:int}
\end{equation} 
where $A(\mathbf{k},\omega)$ describes the one-electron spectral function, $M(\mathbf{k},\omega, t)$ the photoemission matrix element, and $f(\omega, T)$ the Fermi-Dirac (FD) distribution function. Finite experimental resolution is taken into account by convolution with a resolution function $G$, often assumed to be a Gaussian, characterised by $\Delta \mathbf{k}$, $\Delta \omega$, and $\Delta t$. In the most general case, the first three contributions to $I$ can acquire a time dependence: $A$ and $M$ through potential changes in the band structure, and $f$ through the excitation of electrons and the resulting increase in electronic temperature—or indeed through the formation of a transient non-thermal distribution. This general situation, in which all contributions are modified, is not uncommon; it arises, for example, during the melting of a charge density wave or the generation of coherent phonons in the excitation process \cite{Rohwer:2011aa,Sobota:2014aa}.

It is challenging to track such potentially complex behaviour via the measured $I(\mathbf{k},\omega, t)$. A commonly applied technique is to focus on a particular region of interest (ROI) in $(\omega, \mathbf{k})$ space and to extract the intensity integrated over the ROI as a function of time (see, e.g., \cite{Crepaldi:2012ab,Hajlaoui:2012aa,Smallwood:2012aa,Sobota:2012aa,Ulstrup:2014ac,Johannsen:2015aa,Yang:2015ae,Smallwood:2012aa,Majchrzak:2021uw,Bao:2022aa,Zhong:2024aa,Majchrzak:2025aa}). We call this time-dependent intensity in a ROI a ``time distribution curve'' (TDC) \cite{Majchrzak:2025aa}. Such TDCs are typically taken at different locations in $(\omega, \mathbf{k})$ space. They can reveal the decay of an excited electron population or intensity oscillations due to coherent phonon excitation. Due to the potential complexity of the underlying physics, it is generally not trivial to fit the TDCs with a simple model function. Moreover, since TDCs are only extracted at selected $(\omega, \mathbf{k})$ regions, one does not make use of the entire available data set. Recently, it has been proposed to address these limitations by applying $k$-means clustering, an unsupervised machine learning technique, to group TDCs from an entire data set into regions of similar line shape \cite{Majchrzak:2025aa}. This approach can reveal trends in the data that would otherwise go unnoticed. It also has the advantage of not relying on a specific fit model to describe the data and of being easily extendable to data sets in which many experimental parameters are varied.

In this work, we apply $k$-means clustering to TR-ARPES data from quasi-free-standing monolayer graphene (QFMLG) \cite{Riedl:2009aa,Speck:2010aa}, a well-studied model system \cite{Johannsen:2013ab,Gierz:2013aa,Ulstrup:2014aa,Johannsen:2015aa,Na:2019aa,Curcio:2021uh}. This is one of the simplest possible cases in terms of electron dynamics: the band structure near the Fermi energy $E_\mathrm{F}$ consists of the Dirac cone, of which only one branch is observable along the $\Gamma$–K direction in ARPES \cite{Shirley:1995aa,Mucha-Kruczynski:2008aa}. The spectral function and photoemission matrix elements are not expected to change upon pumping, and the only time dependence thus stems from the decay of the excited carrier population. For delay times that are not too short, typically on the order of $>50$~fs, the carrier distribution is well described by a FD distribution with a high electronic temperature. The temperature decay, however, is somewhat complex and requires a description by a three-temperature model, due to the bottleneck for the decay of hot carriers near the Dirac point of the electronic dispersion \cite{Butscher:2007aa,Tse:2009aa,Johannsen:2013aa}.
When revisiting TR-ARPES data from QFMLG using $k$-means, it appears that the decay time of the excited carriers depends not only on $\omega$, as would be expected (since $\omega$ enters the FD distribution), but also on $\mathbf{k}$, which should not influence it. To the best of our knowledge, this apparent $\mathbf{k}$-dependence has not been previously noted in the literature---probably because it is difficult to identify on the basis of established analytical approaches. Employing a simple model, we show that this apparent $\mathbf{k}$-dependence results from the finite energy resolution of the experiment; and the same effect induces an apparent time-dependence of the observed electronic dispersion above $E_\mathrm{F}$. We use our simple model to explore the general factors determining the TDC line shape in TR-ARPES, particularly energy resolution and time resolution. Note that these two are fundamentally linked via the uncertainty principle in TR-ARPES, such that finding a compromise between them—and understanding the influence of both—remains important, even in the presence of experimental improvements \cite{Gauthier:2020aa}.

\section{Methods}

TR-ARPES data were collected from quasi free-standing hydrogen-intercalated graphene on 6H-SiC(0001). The sample was prepared \emph{ex situ}, with a carrier density of approximately $5\times 10^{12}~$cm$^{-2}$ \cite{Johannsen:2013ab} (see Ref.~\cite{Speck:2011aa} for a detailed description of the sample fabrication). Experiments were carried out at the Artemis laboratory of the UK Central Laser Facility. The pump and probe beams, with energies of 1.43~eV and 39.5~eV respectively, were generated from a 100-kHz laser system~\cite{Thire:2023aa}. These beams arrive nearly collinearly at the sample face at an angle of $45^{\circ}$ relative to the detector. Pump and probe beam were polarised perpendicular to and parallel to the plane of incidence, respectively. The pump fluence was 1.3~mJcm$^{-2}$, and the sample temperature was approximately 80~K. The combined energy resolution was $\Delta E = 110$~meV, the momentum resolution was 0.02~\AA$^{-1}$ and the time resolution was $\Delta t =$ 110~fs.

We also simulate TR-ARPES data using a minimal one-dimensional model. We neglect matrix element effects and assume a static spectral function based on a single band with linear dispersion $\epsilon(k) = v k$ and a constant imaginary part of the self-energy $\Sigma''$ as

\begin{equation} A(k,\omega) = \dfrac{1}{\pi}\dfrac{\vert \Sigma'' \vert}{[\hbar \omega - \hbar vk]^2+\Sigma''^2}. \label{equ:specfunc} \end{equation} 
We consider only one direction in $\mathbf{k}$-space, and hence treat $k$ as a scalar quantity.
Specifically, we set the velocity $v = 1.9 \times 10^{6}$~ms$^{-1}$ and $\Sigma'' = 100$~meV. In this model, the only time dependence of $I(k,\omega, t)$ is via the electronic temperature $T(t)$ in the thermal FD distribution:

\begin{equation} f(\omega, T_t) = \dfrac{1}{e^{\hbar \omega / k_BT(t)}+1}. \label{eq:FD} \end{equation}

\noindent We adopt a simple exponential decay of $T(t)$, following an instantaneous excitation at $t = t_0$:
\begin{equation} T(t)=T_0 + (T_{\mathrm{max}} - T_0)H(t - t_0)e^{-t/\tau}, \label{eq:Tdecay} \end{equation} where $\tau$ is a time constant, $T_0$ the equilibrium temperature, $T_{\mathrm{max}}$ the highest electronic temperature reached, and $H(t)$ the Heaviside function, used to describe the instantaneous excitation with perfect time resolution. As model parameters we choose: $T_0 = 80$~K, $T_{\mathrm{max}} = 2500$~K, and $\tau = 500$~fs.
We stress that $\tau$ is the decay time of the temperature, \emph{not} the lifetime of excited states. The latter can only be defined when the decay of $I(t)$ is well described by an exponential, which is not necessarily the case in our model. Moreover, even if it is, the interpretation of the exponential decay time constant in terms of a lifetime may not always be justified \cite{Yang:2015ae,Freericks:2021aa}.
For the simulations including resolution effects, the spectral function is convoluted with a Gaussian along the appropriate dimension (energy or time).

To investigate trends in line shape variation throughout an entire data set, we used $k$-means clustering, an unsupervised machine learning technique. $k$-means groups TDCs into a set of $k$ clusters, where $k$ is defined by the user \cite{MacQueen:1967aa,Ball:1967aa,Xu:2008aa}.
Our approach enables clearer visualisation and analysis of how the TDCs change depending on $\omega$ and $\mathbf{k}$. Importantly, $k$-means is agnostic with respect to the line shape of the TDCs, so no fitting function needs to be defined prior to clustering. The clustering process yields so-called ``cluster centroids,'' which represent the mean TDC within each cluster. These cluster centroids allow us to examine the clustering approach that the algorithm has adopted; and, moreover, they can be treated as averaged, high-signal-to-noise plots of the time dependency within each cluster. 

We apply $k$-means clustering to the TDCs of the entire experimental data set. We first smooth the data in energy and $\mathbf{k}$ using a two-dimensional Gaussian kernel with a full width at half maximum (FWHM) of  $106$~meV and $0.02$~\AA$^{-1}$, respectively.  
We then divide the energy- and $\mathbf{k}$-dependent photoemission intensity into a regular grid of ROIs, from which the TDCs are extracted. These TDCs undergo Gaussian smoothing in time (FWHM=70~fs), followed by subtraction of the pre-time-zero intensity. As we are interested in changes in the TDC line shape with $\omega$ and $\mathbf{k}$—rather than in overall intensity variations across the data set—all TDCs are normalised to a maximum of 1 prior to applying the clustering algorithm. The analysis is based on the code in Ref. \cite{github} accompanying Ref. \cite{Majchrzak:2025aa}, where further details can be found.

\section{Results and Discussion}

We first inspect the experimental and simulated data sets using the conventional approach of analysing TDCs for selected ROIs, establishing a few essential trends in the TDC line shape. We then proceed to a systematic analysis of all experimental and modelled data using $k$-means. This is followed by a more in-depth discussion of the effects resulting from finite time and energy resolution.

\subsubsection{Graphene data and model}

\begin{figure}[h]
    \includegraphics[width=\linewidth]{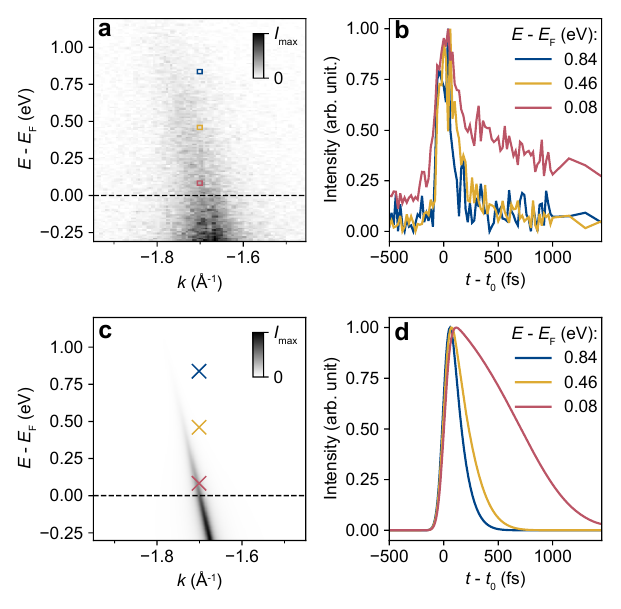}
\caption{(a) Photoemission spectrum from QFMLG at peak excitation  ($t- t_0 $= 20~fs) and $h\nu  =$39.5~eV. (b) Normalized TDCs extracted by integrating the photoemission intensity in the regions of interest marked in (a). (c) Results of the spectral function simulation at peak excitation ($t- t_0 $= 78~fs) and (d) corresponding normalised TDCs extracted at the marked positions (which correspond to those in panel (a)).}
    \label{fig:intro}
\end{figure}

Fig.~\ref{fig:intro}(a) shows the ARPES data from QFMLG near peak excitation ($t - t_0 = 20$~fs). An analysis of the time-dependent photoemission intensity in the three coloured ROIs is presented in panel (b). Note that the ROIs are quite small, leading to a poor signal to noise ratio. They have been chosen like this because we shall use the same ROI size later when clustering all the data. The TDCs inFig.~\ref{fig:intro}(b) have all been normalised to a maximum value of 1 to facilitate comparison of the line shapes. In all three cases, a steep increase in $I(t)$ is followed by a slower decay. The width of the rise is determined by the time resolution and is therefore the same for all three TDCs. The decay time depends on the chosen energy of the ROI: the higher the energy, the faster the decay. This follows trivially from the non-linear form of the FD distribution. However, as we shall see, defining a single ``decay time'' is somewhat problematic, as the TDCs are generally not well described by a single exponential decay. This becomes evident in Fig.~\ref{fig:intro}(b): the TDC closest to $E_\mathrm{F}$ (burgundy) shows more complex behaviour, with an initial fast decay followed by a slower one. While this may appear to reflect the well-known presence of two distinct time scales in the relaxation of the electronic temperature in graphene \cite{Butscher:2007aa,Tse:2009aa,Johannsen:2013aa}, we will show below that the relationship between TDC line shape and electronic temperature decay is too complicated for such a simple description, even in the case of a simple exponential temperature decay.

The corresponding simulations from our simple model are shown in Fig.~\ref{fig:intro}(c) and (d). In panel (c), the spectral function is displayed near peak excitation ($t - t_0 = 78$~fs in this case). The TDCs in panel (d) are taken at the same energies as in the experiment. The ordering of decay times in the TDCs is consistent with the experimental results, with the highest energy showing the fastest decay. What stands out is the complex line shape of the TDC at the lowest energy (burgundy).
TDC line shapes are commonly analysed using exponential decay fits. However, this approach is not always appropriate—especially near the Fermi level. In particular, the observed line shape here cannot be accurately described by a simple (exponential) decay function, despite the fact that the temperature decay in the model is purely exponential.
We will explore the variations in the resulting TDC line shape in greater detail later, focusing on the influence of energy resolution ($\Delta E$) and time resolution ($\Delta t$) on the data.

\subsubsection{$k$-means clustering}

\begin{figure}[ht]
    \includegraphics[width=1\linewidth]{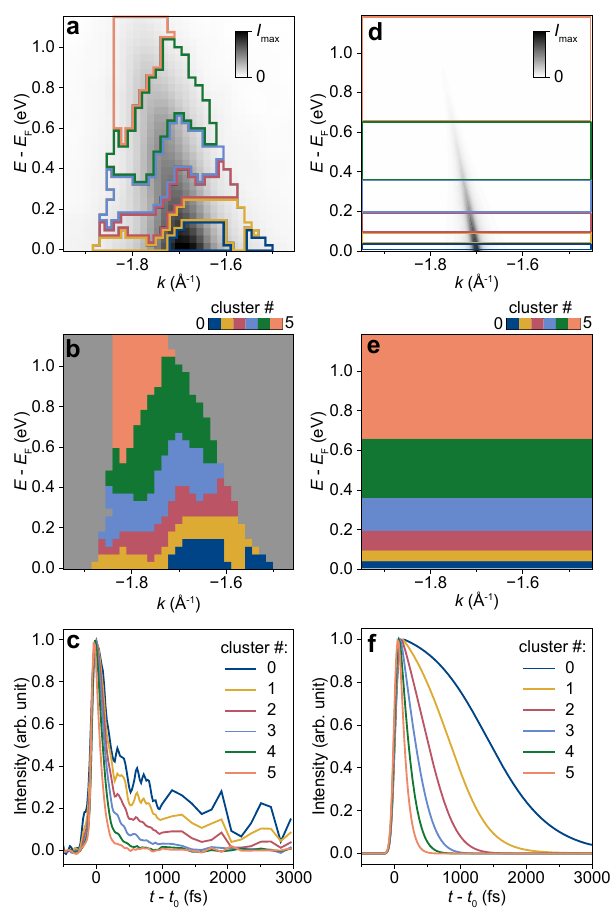}
    \caption{Comparison of $k$-means clustering analysis applied to experimental data (a)-(c) and simulated data without the inclusion of energy resolution effects (d)-(f). (a) Probed photoemission intensity near peak excitation after smoothing the data from Fig. \ref{fig:intro}(a). The coloured lines mark the limits of the areas assigned to one particular cluster resulting from $k$-means. (b) Map of the cluster distribution. The colour assigned to each rectangular ROI indicates that the corresponding TDC belongs to the cluster number indicated in the inset. Grey areas indicate background regions dominated by noise with no well defined TDCs. (c) Cluster centroids. (d)-(f) Corresponding data from the simulation. }
    \label{fig:cluster1}
\end{figure}

The clustering results for the experimental data set are shown in Fig.~\ref{fig:cluster1}(a)–(c). In the clustering process, each ROI is assigned to one of six possible clusters.
Fig.~\ref{fig:cluster1}(a) shows the distribution of these clusters as coloured outlines overlaid on the photoemission intensity near peak excitation after smoothing, highlighting the boundaries between areas assigned to different clusters. Fig.~\ref{fig:cluster1}(b) shows the cluster distribution alone. TDCs from the ROIs in the grey areas in panel (b) were not clustered due to an insufficient signal-to-noise ratio. The fact that the different clusters are well represented by six distinct and almost continuous areas with well-defined and sharp outlines indicates a reliable clustering result, especially considering that the $k$-means algorithm does not take the spatial location of a ROI in $(\omega, \mathbf{k})$ space into account.
Fig.~\ref{fig:cluster1}(c) shows the cluster centroids. As already observed in Fig.~\ref{fig:intro}, a higher average energy of the cluster leads to a faster overall decay. An unexpected result is the $\mathbf{k}$-dependence of the clustering distribution in Fig.~\ref{fig:cluster1}(a),(b). After all, the FD distribution—which is expected to be the only source of time dependence in the photoemission intensity—does not contain any $\mathbf{k}$-dependence. Moreover, the effect is not small: in various regions of $\mathbf{k}$-space (for example, at 0.2~eV above the Fermi level), up to three different clusters can be found with the same energy. These clusters exhibit a strong variation in dynamics (see Fig.~\ref{fig:cluster1}(c)), and this $\mathbf{k}$-dependence could therefore significantly impact any simple analysis based on photoemission intensity in a few arbitrarily selected ROIs.

The corresponding results for the simulated photoemission intensity are shown in Fig.~\ref{fig:cluster1}(d)–(f). These are in stark contrast to the findings from the experiment. First, there is no ${k}$-dependence observed in the clustering of the simulation—the different clusters are separated by horizontal lines, constant in energy. Moreover, the cluster centroids in the simulation exhibit unusual line shapes that clearly deviate from simple exponential decay. In particular, the cluster centroid corresponding to the region closest to the Fermi level (cluster 0, dark blue line in Fig.~\ref{fig:cluster1}(f)) shows a pronounced upward bulging.
Note, however, that this behaviour is found only in a small number of ROIs within a narrow energy range near $E_\mathrm{F}$. Even if present in the experiment, it might go undetected due to finite energy resolution. (A similar but weaker effect is also observed in cluster 1, suggesting that the effective energy range for this behaviour may not be as limited as it first appears.)
Additionally, the simulated data can be clustered even in the regions marked grey in the experimental analysis. This is due to the absence of noise in the simulation, which allows the TDC line shape to be determined even in ROIs where the intensity is several orders of magnitude below the overall peak intensity.

\begin{figure}[ht]
    \includegraphics[width=1\linewidth]{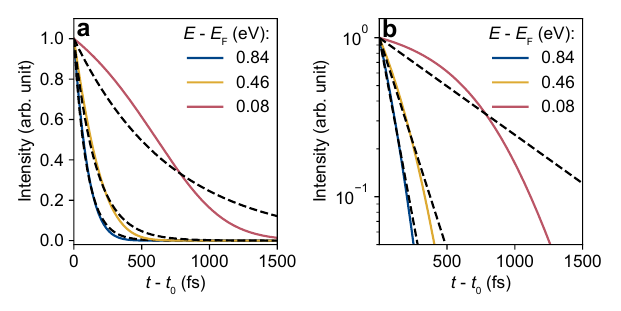}
    \caption{(a) Selected TDCs from the spectral function simulation shown in Fig. \ref{fig:cluster1}(f), but starting at $t - t_0 = 0$, and thus only showing the decay of the TDCs. The decay has been fitted, in each case, using a single exponential decay (dashed line). (b) The same simulations on a log scale,  plotted down to 5\% of the peak intensity. }
    \label{fig:expfit}
\end{figure}

In view of the strong deviation from a single exponential decay—even in this simple model—it is reasonable to ask why fitting with a simple exponential decay is so common and seemingly successful in the analysis of experimental data (graphene being a notable exception \cite{Butscher:2007aa,Tse:2009aa,Johannsen:2013aa}, as pointed out in the introduction). Indeed, when focusing solely on (semi)metallic materials, in order to avoid additional complications from the presence of a band gap, fitting to single exponential decays often appears to work well. In some cases, the fits can be extended to energies close to $E_\mathrm{F}$ \cite{Ulstrup:2014ac,Johannsen:2015aa,Majchrzak:2021uw,Bao:2022aa}, whereas in others, the exponential model fits well only for the initial decay \cite{Smallwood:2012aa}. On the other hand, more complex behaviour with delayed onsets has also been reported \cite{Yang:2015ae}.
A possible explanation for the success of fitting with a single exponential is illustrated in Fig.~\ref{fig:expfit}. Panel (a) shows a subset of the TDCs from Fig.~\ref{fig:cluster1}(f) (excluding the onset and assuming perfect time resolution), along with exponential fits performed under the constraint that the starting value of the intensity equals one at $t = t_0$. Fig.~\ref{fig:expfit}(b) shows the same data on a logarithmic scale, with the lower limit of the scale set to 5\% of the maximum value, to approximate what is typically observable in an experiment in the presence of noise. Both panels show that the blue and yellow simulated TDCs at higher energies are described relatively well by exponential decays, while the burgundy TDC simulated for an energy closer to the Fermi energy diverges significantly from an exponential functional form.

More formally, this can be understood by approximating eq. (\ref{eq:FD}) at high energies—reducing it to a Boltzmann distribution—and eq. (\ref{eq:Tdecay}) at high temperatures, neglecting $T_0$. 
The two expressions are then combined and any non-time-dependent factors are normalised out. If we further assume $t \ll \tau$, we can apply a first-order Taylor expansion to the exponential describing the temperature decay ($e^{t/\tau} \approx 1 + t / \tau$) and express the time-dependent intensity as a simple exponential decay:
\begin{equation} 
I(\omega, t) \propto e^{- \hbar \omega t / \tau k_B T_{\mathrm{max}}}.
 \label{eq:FD2} 
 \end{equation}
While this approximation relies on several assumptions—valid only at high energies and high temperatures, and thus during the early stage of the decay—it provides justification for fitting the initial part of the decay with an exponential function.
Moreover, in the high-energy regime, the decay time is short, and so is the time range over which any deviation from a perfect exponential fit might appear, before the signal reaches the noise level. Finally, the condition $t \ll \tau$ can be easily satisfied, since $\tau$ is the decay time of the temperature in Eq.~(\ref{eq:Tdecay}) and can therefore be much longer than the decay time of a TDC at high energy.
It is interesting to note that the exponential decay described by Eq.~(\ref{eq:FD2}) contains an ``effective decay time" given by $\tau k_B T_{\mathrm{max}} / \hbar \omega$, which would be extracted by fitting to experimental data. However, since neither $T_{\mathrm{max}}$ nor $\tau$ are typically known, this expression cannot be directly used to determine the parameters governing the electronic temperature decay (though $T_{\mathrm{max}}$ can often be determined by other means \cite{Ulstrup:2014aa}).

 
\subsubsection{Finite energy resolution effects}

\begin{figure*}[t]
    \centering
    \includegraphics[width=0.8\linewidth]{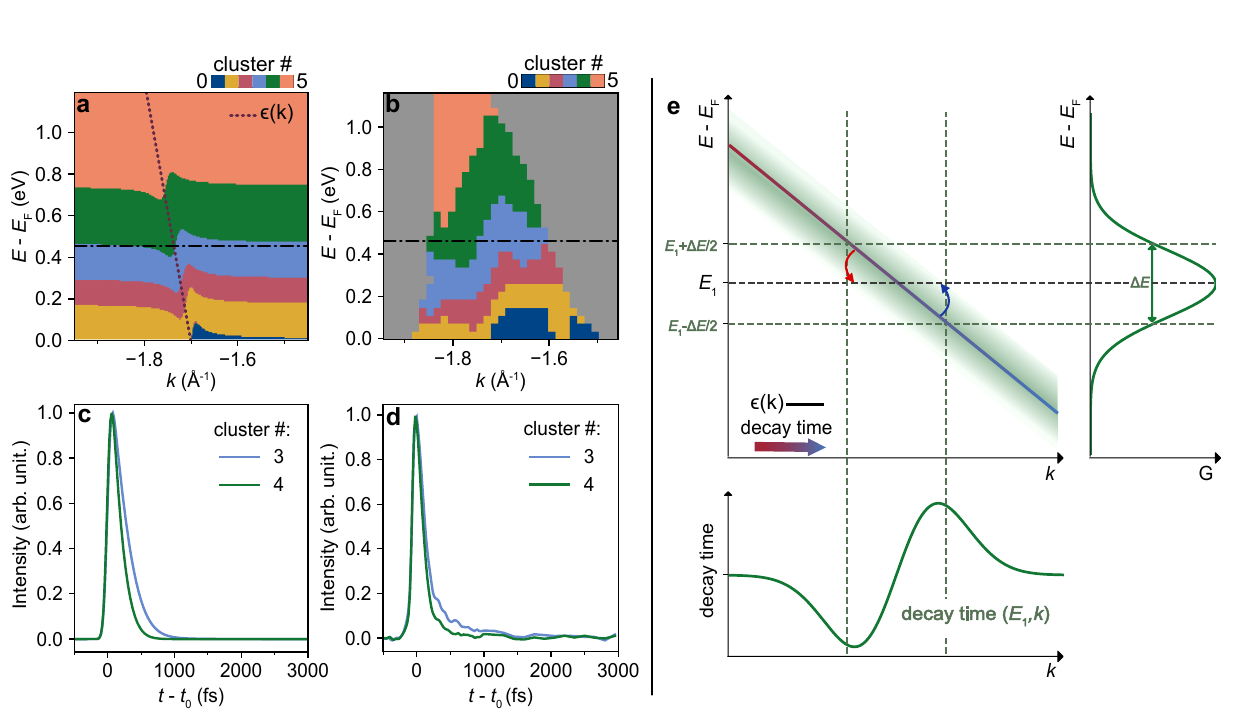}
    \caption{(a,b) Cluster distribution resulting from applying $k$-means  analysis  of (a) simulated data including a finite energy resolution ($\Delta E = 220$~meV) and  (b) experimental data of QFMLG. The bare band $\epsilon(k)$ in the simulated data is indicated by the dotted line.
(c,d) Cluster centroids corresponding to the cluster regions intersected by the dashed-dotted lines at constant energy in panels (a) and (b).
(e) Schematic illustration of the effect causing the $\mathbf{k}$-dependent cluster distribution in the presence of finite energy resolution.  A Gaussian convolution (top right panel) is applied to a linear bare  band (main panel), for which the decay time decreases with increased energy, as shown by a colour gradient from blue to red. The resulting $\mathbf{k}$-dependence of the observed decay time at energy $E_1$, marked in the main panel, is schematically illustrated in the bottom panel.}
    \label{fig:schematic}
\end{figure*}

The wave-shaped cluster distribution observed in the experimental data in Fig.~\ref{fig:cluster1}(b) can be reproduced in the simulation when a finite energy resolution is taken into account. This is shown in Fig.~\ref{fig:schematic}(a), where we present the results of our clustering analysis after applying a Gaussian broadening with $\Delta E = 220$~meV to the simulated data.
The resulting cluster regions now exhibit a variation in $\mathbf{k}$, and a wave-like pattern—similar to that seen in the experimental data—emerges for momenta within approximately $\pm 0.025$~\AA$^{-1}$ of the bare band ($\epsilon(k)$, indicated by a purple dashed line in panel (a)). The variation appears slightly stronger in the experimental data shown in Fig.~\ref{fig:schematic}(b); this is in part due to the tendency of noise-free theoretical calculations to produce very small clusters near $E_\mathrm{F}$, resulting in different cluster boundaries. 

The finite energy resolution thus introduces a $\mathbf{k}$-dependence in the TDC shape, with potentially significant consequences for analyses that do not account for it. In particular, TDCs taken at the same energy but at different $\mathbf{k}$ can exhibit different line shapes. This effect may lead to a broadening of the extracted TDC due to averaging over multiple decay times, and it could contribute to the few reported cases in which a $\mathbf{k}$-dependence of the decay constant or line shape has been observed \cite{Smallwood:2012aa,Zhong:2024aa,Majchrzak:2025aa}.
Figure~\ref{fig:schematic}(c) and (d) illustrate the qualitative agreement between experiment and theory by showing the mean TDCs (cluster centroids) extracted along a representative constant energy line, marked by dashed-dotted lines in Fig.~\ref{fig:schematic}(a) and (b). This particular energy slice was chosen to intersect multiple clusters, resulting in distinctly different decay behaviours among the corresponding cluster centroids—even though the underlying TDCs were all extracted at the same  energy.

The mechanism responsible for the wave-like cluster distribution in the presence of finite energy resolution is explained in Fig.~\ref{fig:schematic}(e). When energy resolution is neglected, every normalised TDC at a given energy (e.g., $E_1$, blue dashed line) is identical. The absolute intensity maximum along the dashed line may vary significantly—typically following the Lorentzian distribution that results from a constant energy cut through the spectral function $A(k, \omega)$ \cite{Hofmann:2009ab}—but this variation is removed by normalisation.
However, when the simulated bare band is convoluted with a Gaussian distribution with a full width at half-maximum of $\Delta E$, intensity from both higher and lower energies contribute to the photoemission intensity at $E_1$, as indicated by the red and blue arrows in the schematic. Because electrons at higher energies (red arrows) decay faster—due to the non-linearity of the FD function (as illustrated by the colour gradient from red to blue in the bare band)—the effective TDC extracted from the left-hand side of the bare band will exhibit a slightly faster decay. Conversely, contributions from lower-energy electrons (blue arrows) on the right-hand side of the bare band will slow the decay. This asymmetry leads to the observed up-down wave pattern in the cluster distribution.
Furthermore, since the intensity strongly increases toward $E_F$, the contribution from lower-lying energy states---and thus the corresponding increase in the decay time of the extracted TDC---is expected to be larger at $\mathbf{k}$ values corresponding to smaller $E$. For simplicity, this latter effect has been omitted in the schematic.
Interestingly, the situation described here is closely related to the well-known relationship between energy resolution and $\mathbf{k}$-resolution  in ARPES: finite $\mathbf{k}$-resolution leads to broadening in energy, and finite energy resolution leads to broadening in $\mathbf{k}$. The coupling between the two depends on the slope of the band \cite{Kevan:1986aa}. Fig.~\ref{fig:schematic}(e) could also serve to illustrate this relation, as the energy width of the shaded band translates into a width in $\mathbf{k}$ along the horizontal axis. The key difference in our case, however, is that, when we introduce time-dependent excitations, the effect is asymmetric: there is a pronounced difference in decay time between states at higher and lower energies.

\begin{figure}[t]
    \centering
    \includegraphics[width=1.0\linewidth]{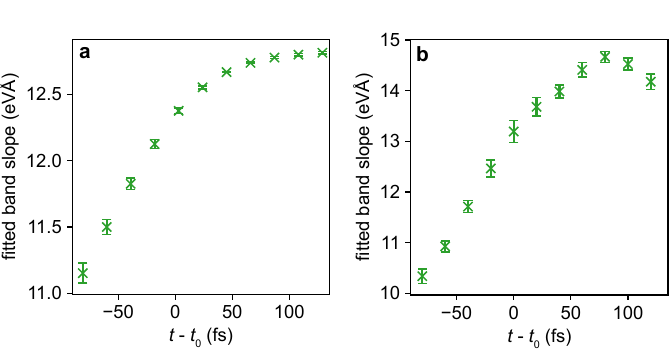}
    \caption{Apparent changes in the of the measured electronic dispersion due to finite energy resolution. The resulting time-dependent slope above $E_\text{F}$ has been extracted from fits to (a) the model data of Fig.~\ref{fig:schematic}(a) and (b) the experimental data shown in  Fig.~\ref{fig:schematic}(b).  }
    \label{fig:dispersion_change}
\end{figure}

Another important consequence of the wave-like cluster distribution in Fig.~\ref{fig:schematic}(a) and (b) is that the apparent dispersion measured by ARPES becomes time dependent. To see this, consider the spectral function in eq. (\ref{equ:specfunc}). When evaluating this at constant energy, the resulting so-called momentum distribution curve (MDC) is a Lorentzian and this is still true when multiplying the spectral function with a Fermi-Dirac distribution. However, a finite energy resolution introduces a variation of the photoemission intensity's time-dependence along the MDCs represented by dashed-dotted lines in in Fig.\ref{fig:schematic}(a) and (b). This implies that the MDC line shape becomes more complex and, in particular, that the mean of this distribution can shift in $\mathbf{k}$, implying a change of the apparent dispersion. The effect is illustrated in Fig. \ref{fig:dispersion_change} where we present the results from fitting a linear dispersion to the MDC maxima in the energy range between $0.006$ and $0.968$~eV for the model data in Fig.\ref{fig:schematic}(a) and for the experimental data in  Fig.\ref{fig:schematic}(b). In both cases, the slope of the band changes considerably as a function of delay time. Considering this resolution effect is important for the interpretation of TR-ARPES data of states above the $E_\mathrm{F}$, as the resulting changes of the apparent dispersion could be mistaken for time-dependent renormalisation effects. 

\subsubsection{Finite time resolution effects}

\begin{figure}[t]
    \centering
    \includegraphics[width=0.5\textwidth]{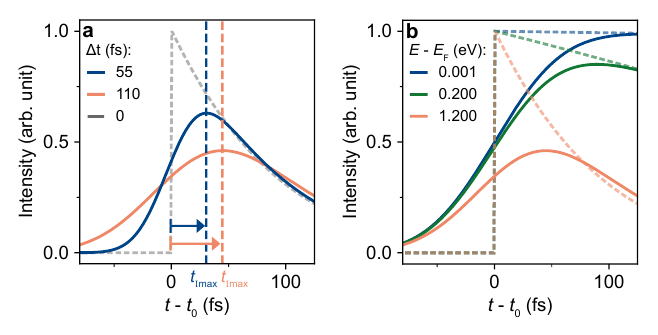}
    \caption{(a) Effect of the energy resolution on the TDC line shape. Dashed line: simulated TDC with perfect energy resolution; solid lines:  TDCs that would be expected for time resolutions of  $\Delta t = 55$ and $\Delta t = 110$~fs, illustrating the broadening and shift in the TDC peak maxima (marked by dashed lines). (b) Effect of the binding energy and time resolution on the TDC line shape. Dashed lines: TDCs with perfect time resolution taken at different energies; solid lines: the same TDCs but for a time resolution of $\Delta t = 110$~fs.  }
    \label{fig:dt}
\end{figure}

The non-negligible effect of $\Delta E$ on the $\mathbf{k}$-dependence and line shape of the TDCs raises the question of whether the time resolution $\Delta t$ might lead to similarly important effects. In the simulations presented so far, a finite time resolution has been included by applying a Gaussian convolution with $\Delta t = 110~$fs. In Fig.~\ref{fig:dt}, we explore the effect of time resolution on simulated spectra. For ease of comparison, $\Delta E$ is set to zero in these simulations.
The dashed line in Fig.~\ref{fig:dt}(a) shows a high-energy TDC—corresponding to a fast decay—calculated with perfect time resolution ($\Delta t = 0$~fs). The solid lines show the same TDC after convolution with different time resolution functions: $\Delta t = 55$~fs (blue) and $\Delta t = 110$~fs (orange). Finite time resolution introduces two important effects. First, it broadens the step at $t = t_0$, as expected. Second, it reduces the peak intensity and shifts the maximum of the TDC to later times. This time shift is on the order of the time resolution, as indicated by the arrows.
Fig.~\ref{fig:dt}(b) illustrates that this shift depends on the TDC energy. When TDCs at different energies—initially calculated with perfect time resolution (dashed lines)—are convoluted with a Gaussian corresponding to $\Delta t = 110$~fs, the resulting time shift decreases at higher energies. Thus, the maximum of the yellow TDC (simulated at 1.2~eV above $E_F$) occurs at a temporal shift of less than 50~fs, while the simulated maximum of the blue TDC (only 0.001~eV above $E_F$) is shifted to more than 100~fs after $t_0$. This is because the un-broadened TDCs at higher energies are already narrower. The relationship between the ideal and broadened curves can be understood by considering two extreme cases: the convolution of a step function (low energy, blue curve in Fig.~\ref{fig:dt}(b) at 0.001~eV) and the convolution of an approximate $\delta$-function (high energy, orange curve at 1.2~eV) with a Gaussian.

Even in the simple model used here, both the magnitude of $\Delta t$ and the binding energy—and thus the decay time of the TDC—contribute to modifications of the TDCs due to the time resolution. Notably, poorer time resolution leads to a more pronounced energy-dependent shift in the position of the TDC maximum. However, these shifts are moderate and generally on the order of the time resolution itself.
In contrast, delayed decay onsets observed in semiconductors or topological insulators are often significantly larger than the time resolution and arise from more complex mechanisms—such as the decay of a surface state population that is simultaneously fed by electrons from the bulk conduction band \cite{Hajlaoui:2012aa,Sobota:2012aa,Wang:2012ad}, or the presence of a time-dependent chemical potential \cite{Crepaldi:2012ab}. Even in high-temperature superconductors, delayed onsets of several hundred femtoseconds have been reported, which clearly cannot be attributed to limited time resolution \cite{Yang:2015ae}.
On the other hand, smaller shifts, such as those reported for PtBi$_2$ \cite{Majchrzak:2025aa}, could plausibly be related to time-resolution broadening.

\section{Conclusion}

In conclusion, we have exploited the potential of $k$-means clustering to analyse both experimental TR-ARPES data from quasi free-standing monolayer graphene and simulations based on a simple linear dispersion. By inspecting the entire data set in this way, we found that the finite energy resolution can significantly influence the observed decay rates, inducing an apparent $\mathbf{k}$-dependence and an apparent time-dependence of the observed electronic dispersion. The time resolution, on the other hand, was found to shift the maximum of a time distribution curve near $t = t_0$, but only to a limited extent—on the order of the time resolution.
Finally, we explored the conditions under which the observed intensity decay can be reasonably approximated by a simple exponential function, as is commonly done in the field. Overall, our results call attention to important subtleties in the interpretation of pump-probe ARPES data, and demonstrate how the application of machine learning to large data sets can reveal underlying trends that might otherwise remain unnoticed.

\begin{acknowledgments}
We acknowledge Florian Speck and Peter Richter for assistance with sample preparation. This project has received funding from the European Union's Horizon 2020 research and innovation programme under grant  agreement no. 871124 Laserlab-Europe. We gratefully acknowledge funding from Independent Research Fund Denmark  (Grant No. 1026-00089B and Grant No. 4258-00002B) and Novo Nordisk Foundation (Grant number NFF23OC0085585).
\end{acknowledgments}

%
%

\end{document}